\documentclass{kluwer}    
\usepackage{epsfig}

\newdisplay{guess}{Conjecture}

\begin{document}
\begin{article}
\begin{opening}
\title{Rms-flux relation of Cyg X-1 with {\it RXTE}: dipping and nondipping cases}
\author{Ya-Juan \surname{Lei}}
\runningauthor{Lei et al.}
\runningtitle{Rms-flux relation of Cyg X-1}
\author{Li-Ming \surname{Song}}
\author{Jin-Lu \surname{Qu}}
\institute{Laboratory for Particle Astrophysics, Institute of
 High Energy Physics, Chinese Academy of Sciences, 100049, Beijing, China;
 leiyj@mail.ihep.ac.cn}
\author{Cheng-Min \surname{Zhang}}
\institute{National Astronomical Observatories, Chinese Academy of Sciences, 100012, Beijing, China;
zhangcm@bao.ac.cn}
\date{Mar 5, 2007}


\begin{abstract}
The rms (root mean square) variability is the parameter for
understanding the emission temporal properties of X-ray binaries
(XRBs) and active galactic nuclei (AGN).
 The rms-flux relation with {\it Rossi X-ray Timing Explorer (RXTE)} data 
 for the dips and nondip of black hole Cyg X-1 has been investigated in this paper.
 Our results show that there exist the linear rms-flux relations in the
frequency range 0.1-10 Hz for the dipping light curve. Moreover,
this linear relation still remains during the nondip regime, but
with the steeper slope than that of the dipping case in the low energy band. 
For the high energy band, the slopes of the dipping and nondipping cases are 
hardly constant within errors.
The explanations of the results have been made by means of the
``Propagating Perturbation'' model of Lyubarskii (1997).
\end{abstract}
\keywords{accretion, accretion disks--binaries: close--circumstellar matter--scattering--
stars: individual (Cygnus X-1)--X-rays: stars}

\end{opening}
\section{Introduction}           

Aperiodic X-ray variability is a general characteristic of XRBs and
AGN (e.g., van der Klis 1994; Vaughan et al. 2003a; McHardy et al.
2004; Uttley et al. 2005). The recent studies show that the linear
rms-flux relation occurs in the X-ray light curves of XRBs and AGN
over a broad range of time-scales, then this linear rms-flux
relation is offset on the flux axis,  which suggests the light
curves to include at least two components: one component with a
linear rms-flux relation and another with a constant rms to flux
(Uttley \& McHardy 2001; Gleissner et al. 2004).
The fact of the same rms-flux relationship   found in the light
curves of XRBs and AGN suggests that this behavior is intrinsic to
the  accreting systems,  and in addition the similar variability
properties  occur, as well,  in neutron star (NS) and black hole
(BH) XRBs (Uttley \& McHardy 2001; Belloni et al. 2002) with the
different X-ray spectra (e.g., Done \& Gierlinski 2003).
Therefore, investigating the  X-ray variability would provide a
probe in tracing the clues of the X-ray emission mechanisms of
compact accreting systems.
%

Various  models have been proposed to explain the shape of
power-spectral density function (PSD) of XRBs, for example, the additive
shot-noise models or flare models where the light curve is produced
by a sum of shots or flares that trigger the avalanches of the
smaller flares (similar to the model of Stern \& Svensson (1996)),
which is caused by the  magnetic reconnection in the corona
(Poutanen \& Fabian 1999).  This variability model is often
mentioned as a ``coronal flare'' model (hereafter CF model, e.g.,
Uttley 2004).
The PSD is a powerful tool to analyze the variability, however it
confronts  some limitations in distinguishing the models for
aperiodic variability. While, these models seek to explain the
observed PSD shapes by the distribution of shot time-scales and shot
profiles which can fit any noise process PSD in theory (see, e.g.,
Miyamoto et al. 1988; Belloni \& Hasinger 1990; Lochner et al. 1991;
Negoro et al. 1995).


As indicated by Uttley \& McHardy (2001),
the linear rms-flux relation found in BH binary Cyg X-1 and NS binary SAX
J1808.4-3658 could provide  the strong  constraints on models of
variability, for example, the simple additive shot-noise or
flare models cannot provide the linear rms-flux relation on short time-scale.
Uttley (2004) demonstrates that the aperiodic variability showing
the linear rms-flux relation in SAX J1808.4-3658 is coupled to the 401 Hz pulsation,
implies this variability originates from the NS surface and not from
the corona, which is most likely associated with the accretion flow.
The ``Propagating Perturbation''  model (PP model for short) of
Lyubarskii (1997) is a promised one, which ascribes the variability
to the variations in accretion rate occurring at different radii
(slow variations occurring at larger radii) while the perturbation
propagates inwards and modulates the energy release in the
X-ray-emitting region and results in the X-ray light curve.
Besides successfully explaining the rms-flux relation, the PP model
can also explain the spectral-timing properties of the variability
 (Kotov et al. 2001; Vaughan et al. 2003a; McHardy et al. 2004).
Similar to the PP model, the exponential model of Uttley et al.
(2005) can infer  the linear rms-flux relation and a lognormal flux
distribution, which is consistently  associated with the
observations.

X-ray dips often occur in the X-ray light curves of XRBs, and are
suggested to be due to the absorption by matter passing through the
line of sight to the X-ray emitting region (White et al. 1995).
Thus, the research on  the dips will provide us some
information  in the X-ray emitting region
(e.g., Shirey et al. 1999; Asai et al. 2000; Church 2001).

In this paper,
using {\it RXTE} data of Cyg X-1, we study statistically the rms-flux relation over
the frequency range 0.1-10 Hz of the dips and nondip, respectively.
In Sect. 2, we introduce the observational data and the
method of calculating rms.
The obtained results on the rms-flux relation are described in Sect. 3.
The discussions about the rms-flux relation, including its slope and intercept,
and models of X-ray variability are given in Sect. 4.

\section{Observation and Data Reduction}


We acquire the public archival data from {\it RXTE} Proportional
Counter Array (PCA) observations of Cyg X-1, and choose the
observations from April 15 of 1996  to  March 22 of 1999 (PCA Epoch
3), which have the same channel-energy relationship.
The observations of PCA Epoch 3 with the data modes
SB\_125us\_0\_13\_2s, SB\_250us\_0\_13\_2s or SB\_500us\_0\_13\_2s
are used. We extract the light curves of every OBSID with the
standard 2 mode over the energy bands of 2-5 keV and 13-24 keV,
where the hardness ratio is defined as the  count rate ratios
between 13-24 keV and 2-5 keV. We find the obvious long term dips
(the duration of all the dips $>$ 1000 s) in the observations
P10241, P30158 and P30161, while in the low state (see Table 1).
The P30161 contains many dips, so, which can be divided into two
parts for analyzing. Figures 1-4 show the light curves and hardness
ratio of the observations containing long term dip, and in the
bottom panels of these four figures (hardness ratio) the dips are
indicated by the arrows.

\begin{table*}
\begin{center}
\caption{\bf ~~The observations containing long term dips and data modes}
\medskip
\begin{tabular}{c c c c c}
\hline
        OBSID       & Time & Data model &   \\
\hline
  P10241     &  96-10-23 18:30:24 to 96-10-25 02:30:24  & SB\_250us\_0\_13\_2s &  \\
\hline
  P30158     &  97-12-14 08:48:14 to 97-12-30 05:30:14  &  SB\_500us\_0\_13\_2s  &  \\

\hline
   30161-01-01-000  & 97-12-28 13:56:00 to 97-12-28 21:03:07  &   SB\_250us\_0\_13\_2s &  \\
   30161-01-01-00   & 97-12-28 21:03:07 to 97-12-29 00:18:14  &  SB\_250us\_0\_13\_2s  & \\
\hline
   30161-01-02-000   & 98-11-13 13:10:33 to 98-11-13 21:10:33  & SB\_250us\_0\_13\_2s &  \\
   30161-01-02-00    & 98-11-13 21:10:33 to 98-11-13 22:00:14  & SB\_250us\_0\_13\_2s & \\
   30161-01-02-01    & 98-11-13 22:40:32 to 98-11-13 23:34:14  & SB\_250us\_0\_13\_2s & \\
\hline
\end{tabular}
\end{center}
\end{table*}

\begin{figure*}
\centerline{\epsfig{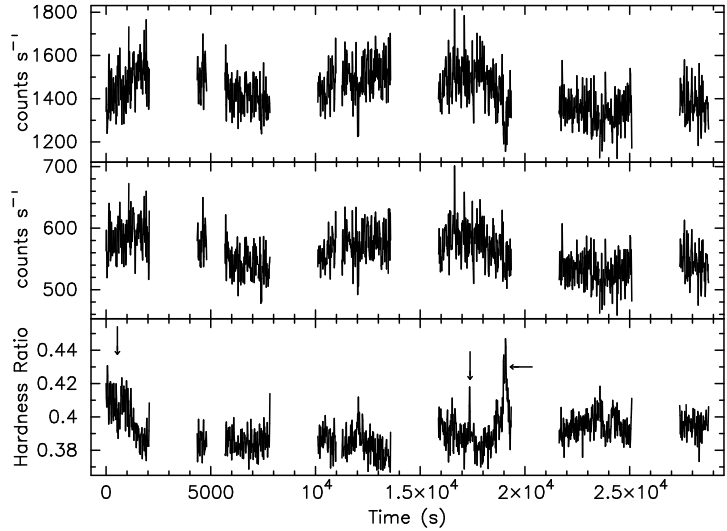}}
\caption{From top to bottom, the light curves of the 2-5
 keV,  13-24 keV,  and the hardness ratio, respectively,  for OBSID P10241.
The hardness ratio is defined as  I(13-24 keV)/I(2-5 keV). Each
point represents 16 s data from all five PCU detectors. The dips are
indicated by the arrows.}
\label{fig1}
\end{figure*}

\begin{figure*}
\centerline{\epsfig{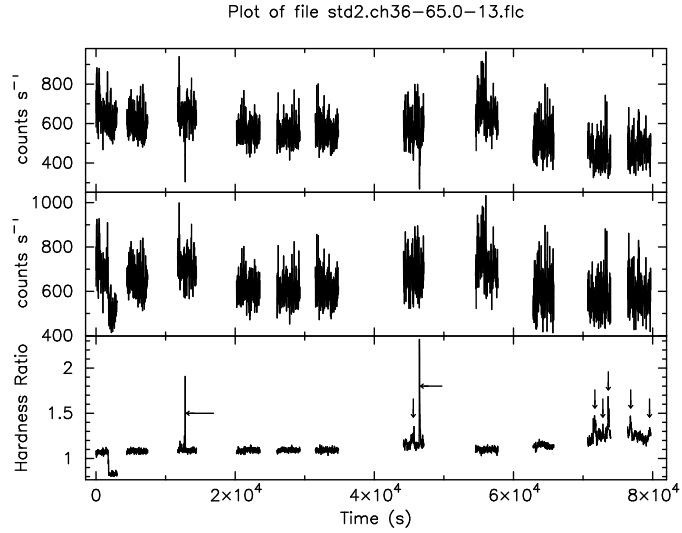}}
\caption{Similar to Fig. 1, but for OBSID P30158.}
\label{fig2}
\end{figure*}

\begin{figure*}
\centerline{\epsfig{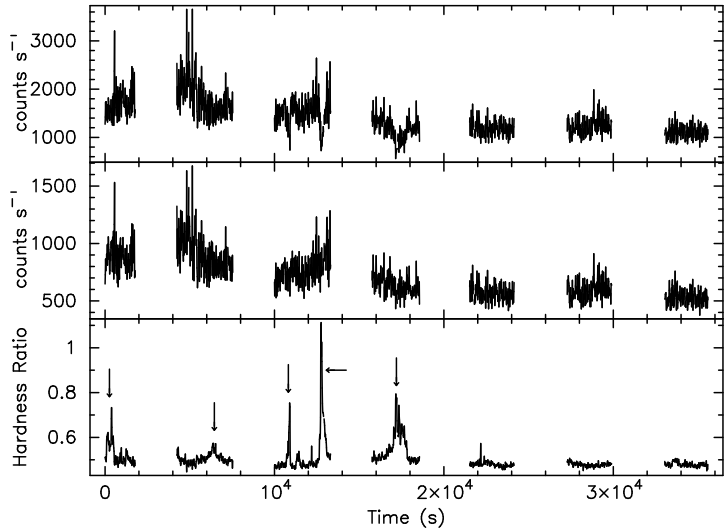}}
\caption{Similar to Fig. 1, but for OBSID 30161-01-01-000 and
30161-01-01-00.}
\label{fig3}
\end{figure*}

\begin{figure*}
\centerline{\epsfig{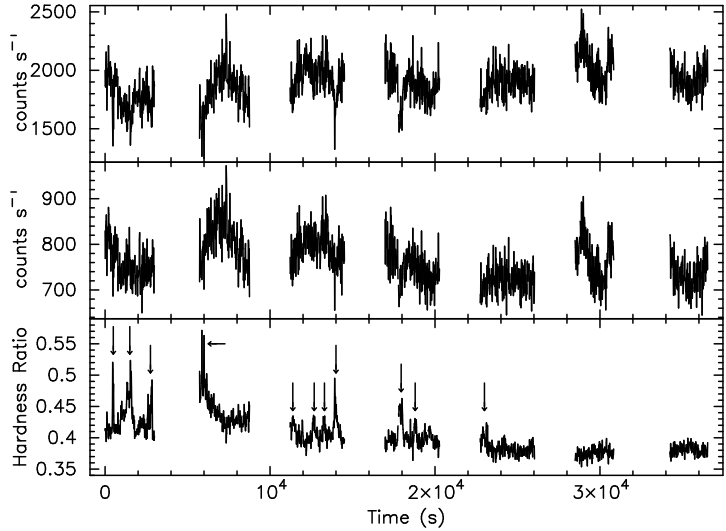}}
\caption{Similar to Fig. 1, but for OBSID 30161-01-02-000,
 30161-01-02-00, 30161-01-02-01.}
\label{fig4}
\end{figure*}

To determine whether the rms variability of a small segment of light
curve depends on the mean flux of that segment for the dips and
nondip, we cut the observed light curve with a resolution of 4 ms
into the segments of 32 s duration for which the source count rate
is determined for computing the rms over the frequency range 0.1-10
Hz, respectively for the energy bands 2-5 keV and 13-24 keV.
These segments are then assigned to their respective flux bins,
and every flux bin contains at least 10 segments.

We adopt the method of Gleissner et al. (2004) to calculate rms, and
it is determined for the  chosen  Fourier frequency interval. We
construct the PSD of the individual light curve segment, applying
the standard rms-squared normalization to the PSD (Miyamoto et al.
1992). We then average all the PSDs obtained from the segments in
the same flux bin, and bin the averaged PSD over the chosen
 frequency range to yield the average power density $\langle P
\rangle$ in that frequency range. Then, the absolute rms variability
of the source, $\sigma$, is given as follows,

\begin{equation}
{\sigma} = {[(\langle P
\rangle-C_{Poisson})\,\Delta\,f]^{1/2}\,F}\;,
\end{equation}
\noindent  where $F$ is the mean count rate of every flux bin.
$C_{\rm Poisson}$ is the Poisson noise level due to photon counting
statistics, and $\Delta f$ is the width of the chosen  frequency
range. The statistical uncertainty of the average PSD value, $\Delta
\langle p \rangle$, is determined from the periodogram  statistics
(van der Klis 1989)

\begin{equation}
\Delta \langle P \rangle = \frac{\langle P \rangle}{\sqrt{MW}}\;,
\end{equation}
\noindent where $M$ is the number of segments used in determining
the average rms value and $W$ is the number of Fourier frequencies.
The uncertainty of $\sigma$, $\Delta$($\sigma$), is then computed
using the standard error propagation formula.

\section{Results}
For the dips and nondip of the studied observations, we analyze
the data of the energy bands 2-5 keV and 13-24 keV, 
corresponding to the data modes. Their rms and flux values are plotted
in Fig. 5-12.
%
We use a function $\sigma$ = $k(F-C)$ to fit the relations, where $F$
is the mean flux of each flux bin and $k$ ($C$)is the slope
(intercept). The linear model provides a good fit to the data.
As shown in Fig. 5-12, 
the fitted linear rms-flux relations
do not pass through the coordinate origins, and there are the
intercepts in the flux axis.
Table 2 shows the values of $k$ and $C$ of all the studied observations.

For the energy band 2-5 keV, the photons are more absorbed by the obscuring matter.
The slopes $k$ of the dips are smaller than that of
nondip for all the studied observations in the energy band,
which implies that the obscuring matter affects the rms variability.
The intercept $C$ of the dips are also smaller than that of
nondip (Table 2), indicating that the constant component is absorbed by 
the obscuring matter.
%
%
%

%
\begin{figure*}
\begin{center}
\centerline{\epsfig{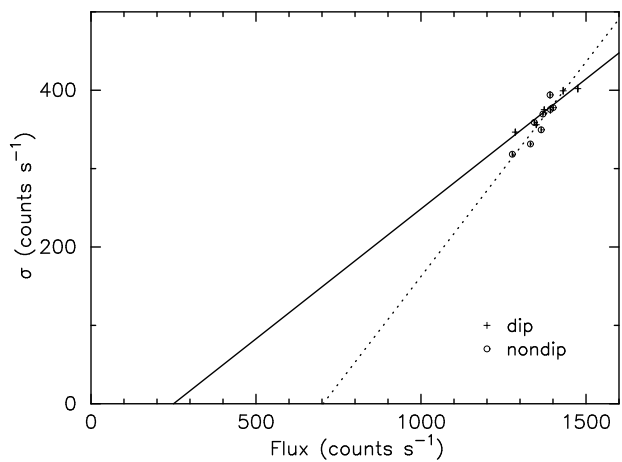}} \caption{Root mean square amplitude $\sigma$
of 0.1-10 Hz over the energy band 2-5 keV vs. X-ray flux for the dips and nondip of
OBSID P10241, where the fitted linear relation $\sigma=k(F-C)$ is
plotted with  $k = 0.33 \pm 0.04$ and  $C = 250 \pm 187$ for the
dips and
 with $k = 0.55 \pm 0.05$ and  $C = 703 \pm 130$ for the nondip,
respectively. The cross (empty circle) stands for the dip (nondip).}
\label{fig5}
\end{center}
\end{figure*}

\begin{figure*}
\begin{center}
\centerline{\epsfig{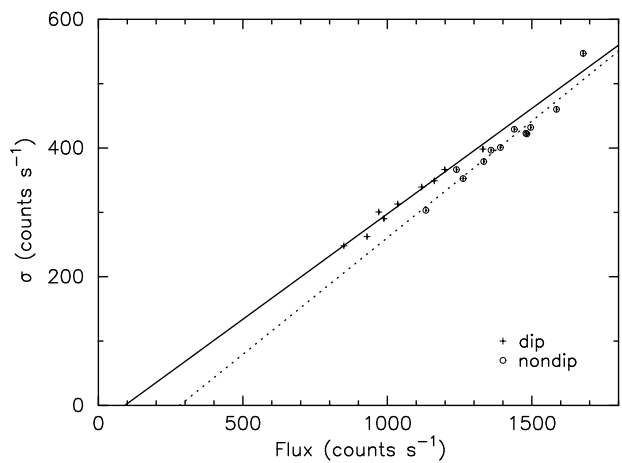}} \caption{Similar to Fig. 5, but for OBSID
P30158, $k = 0.33 \pm 0.01$ and $C = 92 \pm 34$ for the dips, and
 $k = 0.36 \pm 0.01$ and  $C =281 \pm 39$ for the nondip,
respectively.} \label{fig6}
\end{center}
\end{figure*}

\begin{figure*}
\begin{center}
\centerline{\epsfig{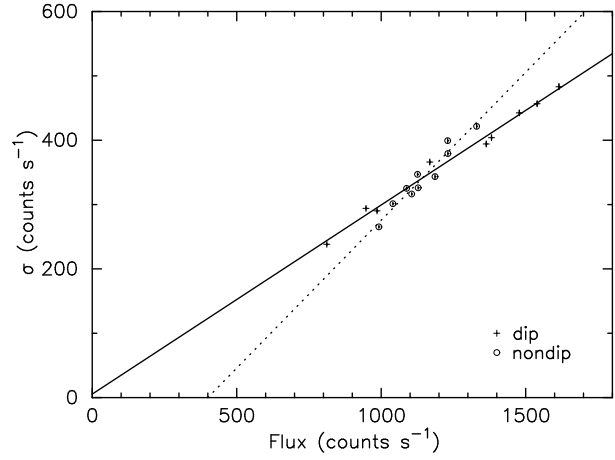}} \caption{ Similar to Fig.5, but for OBSID
30161-01-01-000 and 30161-01-01-00, $k = 0.29 \pm 0.01$ and $C = -20
\pm 24$ for the dips, and
 $k = 0.46 \pm 0.01$ and  $C = 400 \pm 37$ for the nondip,
respectively.} \label{fig7}
\end{center}
\end{figure*}

\begin{figure*}
\begin{center}
\centerline{\epsfig{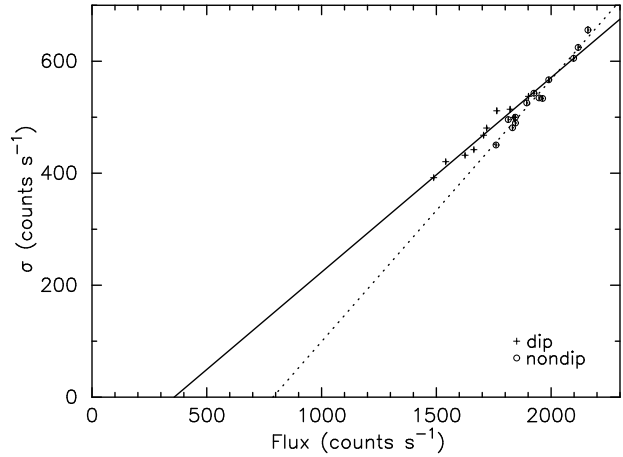}} \caption{Similar to Fig. 5, but for OBSID
30161-01-02-000, 30161-01-02-00 and 30161-01-02-01, $k = 0.35 \pm
0.02$ and $C = 360 \pm 77$ for the dips, and
 $k = 0.47 \pm 0.02$ and $C = 788 \pm 83$ for the nondip,
respectively.} \label{fig8}
\end{center}
\end{figure*}

For the energy band 13-24 keV,
except for the observation 30161-01-01-000 and 30161-01-01-00, 
the slopes $k$ and intercepts $C$ are constant within errors (Table 2), 
which can be due to that the high energy photons is hardly absorbed 
by the obscuring matter.
For the observation 30161-01-01-000 and 30161-01-01-00, 
the obscuring matter could be so compact that affect the high energy photons, 
and lead to the decrease of the slope $k$.
In addition, for the nondip, our results show that the slopes $k$ of 
13-24 keV are smaller than those of 2-5 keV, which 
is consistent with the results of Gleissner et al. (2004) who deduce that 
the slope of the rms-flux relation decreases with the increase of the energy
in the typical hard state observation of Cyg X-1.

Our results show that the obscuring matter does not significantly
influence on the linear rms-flux relation, but reduces the rms
variability of the low energy band. 
In general, the dips are suggested to be due to that the emission regions are 
obscured by the obscuring matter. Therefore, 
the obscuring matter could lead to the decrease of the rms variability.
 %

\begin{table*}
\begin{center}
\caption{\bf ~~The slopes and  intercepts of the linear rms-flux
relations}
\medskip
\begin{tabular}{ c c c c c c}

\hline
 OBSID & dip/nondip &  slope (k) &  intercept (C)& slope (k) &  intercept (C) \\
      &  &\multicolumn{2}{c}{2-5 keV}  &\multicolumn{2}{c}{13-24 keV} \\
\hline
P10241 & dip     & 0.33 $\pm$ 0.04  & 250 $\pm$ 187       & 0.24 $\pm$ 0.05 &12 $\pm$ 144\\
       & nondip  & 0.55 $\pm$ 0.05  & 703 $\pm$ 130       & 0.33 $\pm$ 0.05 & 176 $\pm$ 98\\
\hline
P30158  &dip     & 0.33 $\pm$ 0.01  & 92 $\pm$ 34      &  0.31 $\pm$ 0.02 &13 $\pm$ 100\\
        &nondip  & 0.36 $\pm$ 0.01  & 281 $\pm$  39      & 0.32 $\pm$ 0.01 &80 $\pm$ 19\\
\hline
30161-01-01-000  &dip     & 0.29 $\pm$ 0.01 & --20 $\pm$ 24    & 0.26 $\pm$ 0.01 &--133 $\pm$ 35\\
30161-01-01-00   &nondip  & 0.46 $\pm$ 0.01 & 400 $\pm$ 37  & 0.44 $\pm$ 0.02 & 189 $\pm$ 23 \\
\hline
30161-01-02-000  &dip     & 0.35 $\pm$ 0.02 & 360 $\pm$ 77  & 0.30 $\pm$ 0.03 & 240 $\pm$ 82\\
30161-01-02-00   &        &                     &               &   \\
30161-01-02-01   &nondip  & 0.47 $\pm$ 0.02 & 788 $\pm$ 83  &  0.33 $\pm$ 0.02 & 271 $\pm$ 51\\
\hline
\end{tabular}
\end{center}
\end{table*}

\begin{figure*}
\begin{center}
\centerline{\epsfig{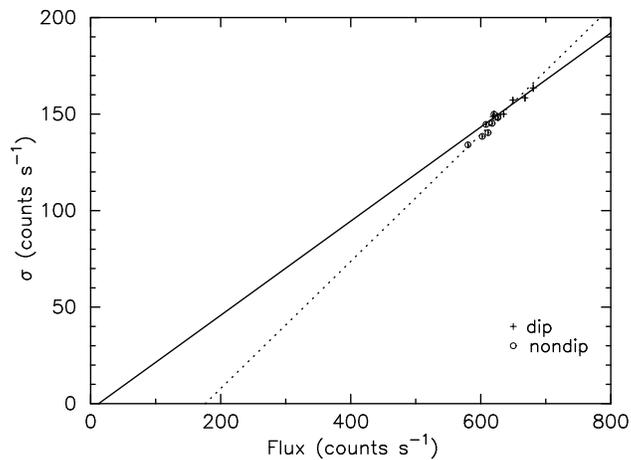}} \caption{Root mean square amplitude $\sigma$
of 0.1-10 Hz over the energy 13-24 keV vs. X-ray flux for the dips and nondip of
OBSID P10241, where the fitted linear relation $\sigma=k(F-C)$ is
plotted with  $k = 0.24 \pm 0.05$ and  $C = 12 \pm 144$ for the dips
and
 with $k = 0.33 \pm 0.05$ and  $C = 176 \pm 98$ for the nondip,
respectively. The cross (empty circle) stands for the dip (nondip).}
\label{fig9}
\end{center}
\end{figure*}

\begin{figure*}
\begin{center}
\centerline{\epsfig{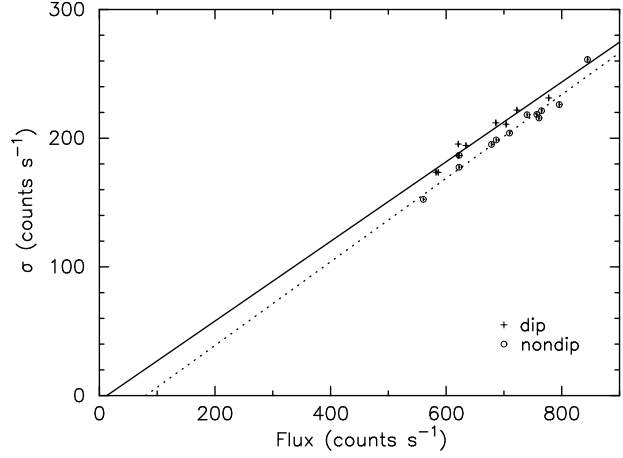}} \caption{Similar to Fig. 9, but for OBSID
P30158, $k = 0.31 \pm 0.02$ and $C = 13 \pm 100$ for the dips, and
 $k = 0.32 \pm 0.01$ and  $C =80 \pm 19$ for the nondip,
respectively.} \label{fig10}
\end{center}
\end{figure*}

\begin{figure*}
\begin{center}
\centerline{\epsfig{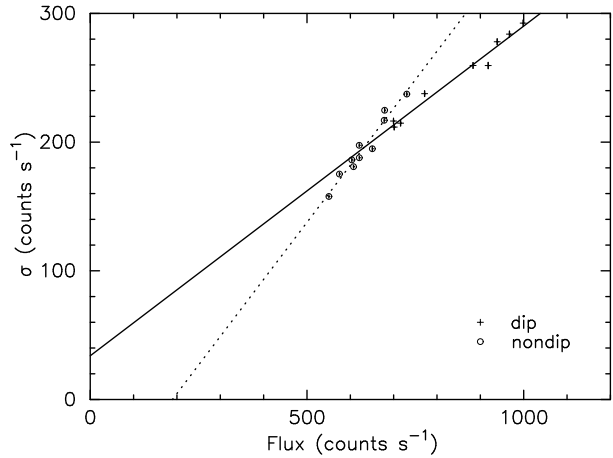}} \caption{ Similar to Fig.9, but for OBSID
30161-01-01-000 and 30161-01-01-00, $k = 0.26 \pm 0.01$ and $C =
-20 \pm 24$ for the dips, and
 $k = 0.44 \pm 0.02$ and  $C = 400 \pm 37$ for the nondip,
respectively.} \label{fig11}
\end{center}
\end{figure*}

\begin{figure*}
\begin{center}
\centerline{\epsfig{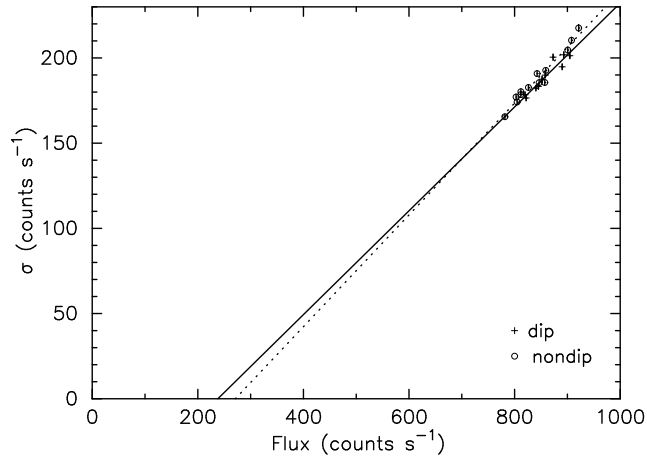}} \caption{Similar to Fig. 9, but for OBSID
30161-01-02-000, 30161-01-02-00 and 30161-01-02-01, $k = 0.30 \pm
0.03$ and $C = 240 \pm 82$ for the dips, and
 $k = 0.33 \pm 0.02$ and $C = 271 \pm 51$ for the nondip,
respectively.} \label{fig12}
\end{center}
\end{figure*}

\section{Discussion}
\subsection{Comparison  with the results of {\it ASCA} observation}

Ba{\l}uci{\' n}ska-Church et al. (1997) study the rms-flux relations
 in the energy bands 0.7-4 keV and 4-10 keV of Cyg X-1 with {\it ASCA},
 and find that
there are (no) dips over the 0.7-4 keV (4-10 keV) energy band
(see Fig.1 of Ba{\l}uci{\' n}ska-Church et al. (1997)).
 Our results show that, for some observations, the dips exist in both 2-5 keV 
 and 13-24 keV bands and the photons of the high energy band
 are also absorbed by the obscuring matter (Fig.1-4), which are not
consistent with the results of Ba{\l}uci{\' n}ska-Church et al.
(1997)who suggest that the X-ray emission region is not abscured in the
high energy band.

The results from {\it ASCA} show that the rms variability is
linearly related with the flux in the 0.7-4 keV band, then  the
rms variability is constant with the flux in the 4-10 keV band.
 Our results from {\it RXTE} show that the rms variability is linearly
 related with the flux in both 2-5 keV and
 13-24 keV bands although both {\it ASCA} and {\it RXTE} observations
 are involved in the hard state of Cyg X-1.

The count rates of nondip data with {\it ASCA} are about 200 count s$^{-1}$
in the 0.7-4 keV band, which are lower than the count rates
from {\it RXTE} in the 2-5 keV band (also see, Fig. 1-4).
Moreover, it is noticed that  the count rates are as low as 60
count s$^{-1}$ in the 4-10 keV band with {\it ASCA},  therefore the
results by Ba{\l}uci{\' n}ska-Church et al. (1997) may be not
conclusive.

\subsection{The rms-flux relation and models of X-ray variability}

The linear rms-flux relations are observed in both XRBs and AGN,
implying that the similar physical mechanisms work in these
accreting systems. No doubt, the linear rms-flux relation provides
the strong constraints on the  models of the X-ray variability.
Uttley \& McHardy (2001) argue that a linear rms-flux relation in
XRBs and AGN gets rid of the additive shot-noise models, since these
models  treat the shots on all time-scales to be independent each
other and cannot predict a linear rms-flux relation in a short time-scale.
After analyzing the rms-flux relation of SAX J1808.4-3658, Uttley
(2004) confirms that the variability could come from the accretion
flow, not the accretion disk corona. In addition, the energy
dependence characteristic of the variability
 disagrees with the conclusions of existing the
 extend corona (Maccarone et al. 2000;
Focke et al. 2005). Furthermore, SOC (self-organized criticality in
accretion flow) is also ruled out because it cannot produce the
observed lognormal flux distribution (Uttley et al. 2005).
Uttley et al. (2005) deduce that PP model seems to be a promising
  model at the present stage. In this model, the
perturbations are treated to be produced at different radii in the
accretion flow, and these perturbations can propagate inwards
without being suppressed and  then the perturbations at different
radii can couple together.
As pointed out, what could fulfill the possibility is a geometrically
thick accretion flow where the accretion time-scales are relatively
short so that the variations can propagate to the inner regions of
the accretion flow without being significantly damped (Manmoto et
al. 1996; Churazov et al. 2001; Ar{\'e}valo \& Uttley 2006).
However, the origin of perturbation is not yet clear, although a
variety of accretion instabilities could contribute to it (Frank et
al. 1992). King et al. (2004) have shown that the
 magnetohydrodynamic turbulence can cause the accretion perturbations on
the sufficiently long time-scales. 

In the PP model, the X-ray variability of low (high) energy band is
produced mostly in the outside (inside) of the accretion flow
(Lyubarskii 1997). The phenomenological exponential  model proposed
by  Uttley et al. (2005), which is mainly consistent with the
conclusions of  PP model, could reproduce the variability properties
of the observed data. Our results show that the slope
of the rms-flux relation decreases with the increase of the energy
in the hard state of Cyg X-1 (also see, Gleissner et al. 2004). The PP model can
provide an appropriate explanation: the photons of high energy band are
produced in the inner accretion regime.
The perturbations in the different outer radii of accretion flow
propagate to the inner accretion flow and modulate the emission
region, as a result, the amplitude of X-ray variability of
high energy will increase.
In fact, the perturbations from the different outer radii to the
inner accretion flow could become stochastic in some extend.
The increase of the stochastic component could weaken the variability of the inner
X-ray emission region, which may result in the decrease of the rms
variability in the high energy band.

%

Our results show that the values of $k$ and $C$ decrease from the
nondip to dip in energy band 2-5 keV (see Table 2), suggesting the
optical thick obscuring matter locate outside of the emission
region. In general, it is believed that the dips of Cyg X-1 are
caused by the density enhancement in an inhomogeneous wind of the
companion star or due to the partial obscuration of an extended
X-ray emitting region  by the optically thick ``clumps'' in the
accretion flow (Feng \& Cui 2002).
%
Therefore, the reason for  $C$ decreasing could be due to the
absorption of the obscuring matter to the constant component.
%
On the other hand, the obscuring matter seems to add a little
stochastic modulation to the intrinsic light curves, which could
make the light curves a bit stochastic and then induce the decrease
of $k$.
%

\subsection{Changes of slope and intercept}

The intercept $C$ of the linear rms-flux relation sometimes is
negative (also see,Gleissner et al. 2004). Therefore, $C$ cannot simply
represent a constant component to the light curve, but can represent
a component with the constant rms. The slope $k$ corresponds to an
increase in the fractional rms of the linear rms-flux component of
the light curve. For convenience, equivalently, the equation $\sigma
= k(F-C)$ can be written as
\begin{equation}
{\sigma = kF+\sigma_{0}}\;,
\end{equation}
where $\sigma_{0}=-kC$ is positive whenever $C<0$, predicting
 $\sigma \to \sigma_{0}$ for $F \to 0$, which
is unphysical. We adopt the assumption that $\sigma \to 0$ for $F
\to 0$, which suggests that, when $\sigma < \sigma_{0}$, other
variability process that does not obey the linear rms-flux
relationship would be important and make $\sigma \to 0$ when $F \to
0$ (also see, Gleissner et al. 2004). When $F$ is from $\simeq 0$ to
a certain flux, suggesting that the value of $k$ increase with the
flux. When $F$ is larger than a certain  flux, the value of $k$ is
up to a constant, or the rms-flux becomes linear.


The PP model confirms that the X-ray  variations produce at the 
different  radii in the accretion flow, with the slower variations 
produced at larger radii. The accretion variations can then propagate
to the small radii, then the variations at different radii,
i.e., different time-scales, can
couple together.
In the emission region of high energy, the variations originate from the
perturbations propagated from the different outer radii of
accretion flow, and they couple with each other,
which induces the increase of the stochastic component and 
the decrease of rms variability.
%
However, the emission region of low energy
is only modulated by the outermost variations,
resulting in that the amplitude of variability
is increasing with flux. 
Therefore, the slope of the high energy band is smaller that that of the low energy band.

\section*{Acknowledgments}
This research has made use of data obtained through the
high energy Astrophysics Science Archive Research Center
Online Service, provided by the NASA/Goddard Space
Flight Center. We acknowledge the RXTE data teams at
NASA/GSFC for their help. This work is subsidized by the
Special Funds for Major State Basic Research Projects and
by the National Natural Science Foundation of China.

\end{article}

\begin{thebibliography}{}


\bibitem[Ar{\'e}valo \& Uttley(2006)]{2006Mon. Not. Roy. Astron. Soc.367..801A} Ar{\'e}valo, P.,
\& Uttley, P.:Investigating a fluctuating-accretion model for the spectral-timing 
properties of accreting black hole systems. Mon. Not. Roy. Astron. Soc 367, 801-804 (2006)


\bibitem[Asai et al. (2000)]{2000ApJS...131.571}
Asai, K., Dotani, T., Nagase, F., \& Mitsuda, K.:Iron K Emission Lines in the 
Energy Spectra of Low-Mass X-Ray Binaries Observed with ASCA. ApJS 131, 571-591 (2000)

\bibitem[Balucinska-Church et al.(1997)]{1997ApJ...480L.115B}
Balucinska-Church, M., Takahashi, T., Ueda, Y., Church, M.~J., Dotani, T.,
Mitsuda, K., \& Inoue, H.: The Cessation of Flickering during Dips in Cygnus X-1. 
Astrophys. J. 480, L115-L119 (1997)

\bibitem[Belloni \& Hasinger(1990)]{1990A&A...227L..33B} Belloni, T., \&
Hasinger, G.:Variability in the noise properties of Cygnus X-1. Astron. Astrophys. 227, L33-L36 (1990)
\bibitem[Belloni et al.(2002)]{2002ApJ...572..392B} Belloni, T., Psaltis,
D., \& van der Klis, M.:A Unified Description of the Timing Features of Accreting X-Ray Binaries. 
ApJ. 572, 392-406 (2002)
\bibitem[Churazov et al.(2001)]{2001Mon. Not. Roy. Astron. Soc.321..759C}
Churazov, E., Gilfanov, M., \& Revnivtsev, M.: Soft state of Cygnus X-1: stable disc and unstable 
corona. Mon. Not. Roy. Astron. Soc 321, 759-766 (2001)
\bibitem[Church(2001)]{2001AdSpR..28..323C} 
Church, M.~J.: The emission regions in X-ray binaries: dipping as a diagnostic. 
Advances in Space Research. 28, 323-335 (2001)

\bibitem[Done \& Gierli{\'n}ski(2003)]{2003Mon. Not. Roy. Astron. Soc.342.1041D}
Done, C., \& Gierli{\'n}ski, M.: Observing the effects of the event horizon in black holes. 
Mon. Not. Roy. Astron. Soc 342, 1041-1055 (2003)

\bibitem[Feng \& Cui(2002)]{2002ApJ...564..953F}
Feng, Y.~X., \& Cui, W.: Discovery of Two Types of X-Ray Dips in Cygnus X-1.  Astrophys. J. 564, 953-961 (2002)
\bibitem[Focke et al.(2005)]{2005ApJ...633.1085F}
Focke, W.~B., Wai, L.~L., \& Swank, J.~H.: Time Domain Studies of X-Ray Shot Noise in 
Cygnus X-1. Astrophys. J. 633, 1085-1094 (2005)

\bibitem[]{}
Frank, J., King, A. R., Raine, D. J.: Accretion Power in Astrophysics, 
Cambridge (1992)

\bibitem[Gleissner et al.(2004)]{2004A&A...414.1091G}
Gleissner, T., Wilms, J., Pottschmidt, K., Uttley, P., 
Nowak, M.~A., \& Staubert, R.: 
Long term variability of Cyg X-1. II. The rms-flux relation. Astron. Astrophys.
414, 1091-1104 (2004)
\bibitem[King et al.(2004)]{2004Mon. Not. Roy. Astron. Soc.348..111K} King, A.~R., Pringle,
J.~E., West, R.~G., \& Livio, M.: Variability in black hole accretion discs. 
Mon. Not. Roy. Astron. Soc. 348, 111-122 (2004)

\bibitem[Kotov et al.(2001)]{2001Mon. Not. Roy. Astron. Soc.327..799K}
Kotov, O., Churazov, E.,
\& Gilfanov, M.: 
On the X-ray time-lags in the black hole candidates. Mon. Not. Roy. Astron. Soc. 327, 799-807 (2001)
\bibitem[Lochner et al.(1991)]{1991ApJ...376..295L} Lochner, J.~C., Swank,
J.~H., \& Szymkowiak, A.~E.: Shot model parameters for Cygnus X-1 through phase portrait fitting. 
Astrophys. J. 376, 295-311 (1991)
\bibitem[Lyubarskii(1997)]{1997Mon. Not. Roy. Astron. Soc.292..679L}
Lyubarskii, Y.~E.: Flicker noise in accretion discs.
Mon. Not. Roy. Astron. Soc. 292, 679-685 (1997)
\bibitem[Poutanen \& Fabian(1999)]{1999Mon. Not. Roy. Astron. Soc.306L..31P}
Poutanen, J., \&
Fabian, A.~C.: Spectral evolution of magnetic flares and time lags in accreting 
black hole sources. Mon. Not. Roy. Astron. Soc. 306, L31-L37 (1999)


\bibitem[Maccarone et al.(2000)]{2000ApJ...537L.107M} Maccarone, T.~J.,
Coppi, P.~S., \& Poutanen, J.: Time Domain Analysis of Variability in Cygnus X-1: 
Constraints on the Emission Models. Astrophys. J. 537, L107-L110 (2000)


\bibitem[Manmoto et al.(1996)]{1996ApJ...464L.135M}
Manmoto, T., Takeuchi, M., Mineshige, S., Matsumoto, R., 
\& Negoro, H.: X-Ray Fluctuations from Locally Unstable 
Advection-dominated Disks. Astrophys. J. 464, L135-L138 (1996)
\bibitem[McHardy et al.(2004)]{2004Mon. Not. Roy. Astron. Soc.348..783M}
McHardy, I.~M., Papadakis, I.~E., Uttley, P., Page, M.~J., 
\& Mason, K.~O.: Combined long and short time-scale X-ray variability of 
NGC 4051 with RXTE and XMM-Newton. Mon. Not. Roy. Astron. Soc.
348, 783-801 (2004)
\bibitem[Miyamoto et al.(1992)]{1992ApJ...391L..21M}
Miyamoto, S., Kitamoto, S., Iga, S., Negoro, H., 
\& Terada, K.: Canonical time variations of X-rays from black hole candidates 
in the low-intensity state. Astrophys. J. 391, L21-L24 (1992)
\bibitem[Miyamoto et al.(1988)]{1988Natur.336..450M} Miyamoto, S.,
Kitamoto, S., Mitsuda, K., \& Dotani, T.: Delayed hard X-rays from Cygnus X-1. 
Nature 336, 450-452 (1988)
\bibitem[Negoro et al.(1995)]{1995ApJ...452L..49N} Negoro, H., Kitamoto,
S., Takeuchi, M., \& Mineshige, S.: Statistics of X-Ray Fluctuations from Cygnus X-1: 
Reservoirs in the Disk?. Astrophys. J. 452, L49-L52 (1995)

\bibitem[Shirey et al.(1999)]{1999ApJ...524.1048S} Shirey, R.~E., Levine,
A.~M., \& Bradt, H.~V.: Scattering and Iron Fluorescence Revealed during 
Absorption Dips in Circinus X-1. ApJ. 524, 1048-1058 (1999)
\bibitem[Stern \& Svensson(1996)]{1996ApJ...469L.109S} Stern, B.~E., \&
Svensson, R.: Evidence for "Chain Reaction" in the Time Profiles of Gamma-Ray Bursts. 
Astrophys. J. 469, L109-L113 (1996)
\bibitem[Uttley(2004)]{2004Mon. Not. Roy. Astron. Soc.347L..61U}
Uttley, P.: SAX J1808.4-3658 and the origin of X-ray variability in X-ray binaries and 
active galactic nuclei. Mon. Not. Roy. Astron. Soc. 347,
L61-L65 (2004)
\bibitem[Uttley \& McHardy(2001)]{2001Mon. Not. Roy. Astron. Soc.323L..26U}
Uttley, P., \&
McHardy, I.~M.: The flux-dependent amplitude of broadband noise variability in X-ray binaries 
and active galaxies. Mon. Not. Roy. Astron. Soc. 323, L26-L30 (2001)
\bibitem[Uttley et al.(2005)]{2005Mon. Not. Roy. Astron. Soc.359..345U}
Uttley, P., McHardy,
I.~M., \& Vaughan, S.: Non-linear X-ray variability in X-ray binaries and active 
galaxies. Mon. Not. Roy. Astron. Soc. 359, 345-362 (2005)

\bibitem[{{van der Klis}(1989)}]{klis:89a}
{van der Klis}, M.: Timing Neutron Stars, ed. H.~\"Ogelman \& E.~P.~J.
  {van den Heuvel}, NATO ASI No. C262 (Dordrecht: Kluwer Academic Publishers),
  27 (1989)

\bibitem[van der Klis(1994)]{1994ApJS...92..511V} 
van der Klis, M.: Similarities in neutron star and black hole accretion. ApJS 92, 511-519 (1994)
\bibitem[Vaughan et al.(2003a)]{2003Mon. Not. Roy. Astron. Soc.339.1237V}
Vaughan, S., Fabian,
A.~C., \& Nandra, K.: X-ray continuum variability of MCG-6-30-15. Mon. Not. Roy. Astron. Soc. 339, 1237-1255 (2003a)
\bibitem[White et al. (1995)]{1995}
White, N., Nagase, F., \& Parmar, A.~N.: X-ray Binaries, ed. W.~G.~H.
Lewin, J. van Paradijs, \& E.~P.~J. van den Heuvel (Cambridge: Cambridge Univ. Press),1 (1995)
\end{thebibliography}
\end{document}